\documentclass[doublecol]{epl2}

\usepackage{epsfig,amsmath,graphicx,graphics,float}

\renewcommand\d[0]{\mathrm{d}}
\newcommand\fixedpartiald[3]{\left.\frac{\partial #1}{\partial #2}\right|_{#3}}

\title{Hybridization of first and second sound in a weakly-interacting Bose gas}
\shorttitle{Hybridization of first and second sound in a weakly-interacting Bose gas}

\author{Lucas Verney\inst{1,2} \and Lev Pitaevskii\inst{1,3} \and Sandro Stringari\inst{1}}

\institute{
    \inst{1} \'{E}cole normale sup\'{e}rieure, International Center for Fundamental Physics, Paris, France\\
    \inst{2} INO-CNR BEC Center and Dipartimento di Fisica, Universit\`a di Trento, I-38123 Povo, Italy \\
    \inst{3} Kapitza Institute for Physical Problems, Kosygina 2, 119334 Moscow, Russia
}


\date{\today}

\abstract{Using Landau's theory of two-fluid hydrodynamics we investigate the sound modes propagating in a uniform weakly-interacting superfluid Bose gas for values of temperature, up to the critical point. In order to evaluate the relevant thermodynamic functions needed to solve the hydrodynamic equations, including the temperature dependence of the superfluid density, we use Bogoliubov theory at low temperatures and the results of a perturbative approach based on Beliaev diagrammatic technique at higher temperatures. Special focus is given on the hybridization phenomenon between first and second sound which occurs at low temperatures of the order of the interaction energy and we discuss explicitly the behavior of the two sound velocities near the hybridization point. }

\begin{document}
\maketitle

Second sound is one of the most interesting manifestations of superfluidity. It
is associated with the motion of the fluid where its normal and superfluid
components oscillate in opposite phase (for a recent discussion see
\cite{Donnelly}). In superfluid Helium, where it was first measured by Peshkov~\cite{Peshkov}, second sound is usually regarded as a temperature
(or entropy) wave and is distinct from first sound which is basically a density
wave. A major interest of second sound is that it gives direct access to the
superfluid density~\cite{Landau47}, a quantity of high physical interest,
vanishing at the superfluid critical temperature. Recently the measurement
of second sound has been used to determine experimentally the temperature
dependence of the superfluid density in a strongly interacting atomic Fermi gas
\cite{Nature}. From a theoretical point of view the description of first and
second sound relies on Landau's two-fluid hydrodynamic equations of superfluidity~\cite{Landau}.
Below the critical temperature, with the appearance of a
superfluid component in the system, the macroscopic dynamics of the system is
described by coupled equations involving the normal and the superfluid
components. The peculiar nature of these equations arises on the one hand from
the hydrodynamic nature of the superfluid motion, strictly related to the
phase of the order parameter, and on the other hand from the collisional regime
characterizing the normal component. In the linearized limit of small
amplitude oscillations, the two-fluid hydrodynamic equations take the form

\begin{equation}
    \frac{\partial^2 \rho}{\partial t^2} = \mathbf{\nabla}^2 P
    \label{HD1}
\end{equation}

\begin{equation}
    \frac{\partial^2 \tilde{s}}{\partial t^2} = \frac{\rho_s \tilde{s}^2}{\rho_n}\mathbf{\nabla}^2 T
    \label{HD2}
\end{equation}
where $\rho_s$ and $\rho_n$ ($\rho_s+\rho_n=\rho=mn$) are the densities of the
superfluid and normal components, $\tilde{s}$ is the entropy density per unit
mass and $P$ is the pressure of the gas.

From Eqs.~(\ref{HD1}-\ref{HD2}), one derives the following quartic equation for
the sound velocities (see for example~\cite{LevSandro})
\begin{equation}
    c^4 - \left( \fixedpartiald{P}{\rho}{\tilde{s}} + \frac{\rho_s T \tilde{s}^2}{\rho_n \tilde{c}_v}\right) c^2 + \frac{\rho_s T \tilde{s}^2}{\rho_n \tilde{c}_v} \fixedpartiald{P}{\rho}{T} = 0
    \label{c4}
\end{equation}
where $\tilde{c}_v$ is the specific heat per unit mass and ${(\partial P / \partial \rho)_T}$ and $(\partial P / \partial \rho)_{\tilde s}$ are the inverse isothermal and adiabatic compressibilities respectively.
Equation (\ref{c4}) admits simple exact solutions at low $T$ and close to the critical temperature. In the first case, where all the thermodynamic functions are fixed by the thermal excitation of phonons, one can easily find the Landau's result ${c^2_2=c^2_1/3}$ where $c_1=\sqrt{(\partial P/\partial \rho)_{T=0}}$ is here the $T=0$ first sound velocity. Close to the critical temperature, where the superfluid density vanishes in 3D systems, the second sound velocity instead tends to zero. In this limit
one can neglect, in Eq.~(\ref{c4}), the small $c^{4}$ term as well as the second term,
proportional to $\rho _{s},$ in the parenthesis multiplying $c^2$. Then, using the thermodynamic
relation $\tilde{c}_{p}/\tilde{c}_{v}=(\partial P/\partial \rho )_{\tilde{s}%
}/(\partial P/\partial \rho )_{T}$, one gets the result:
\begin{equation}
c_{2}^{2}=\frac{\rho _{s}T\tilde{s}^{2}}{\rho _{n}\tilde{c}_{P}}\ .
\label{c2constantP}
\end{equation}
In the same limit the first sound velocity approaches the usual expression
\begin{equation}
c_{1}^{2}=(\frac{\partial P}{\partial \rho})_{\tilde{s}} \ .
\label{c1adiabatic}
\end{equation}

In systems characterized by a small value of the thermal expansion coefficient, such as liquid $^4$He, the adiabatic and isothermal compressibilities are nearly the same and equation (\ref{c4}) admits the very simple solutions ${c^2_1=(\partial P / \partial \rho)}$ and ${c^2_2= (\rho_s T \tilde{s}^2)/(\rho_n \tilde{c})}$ holding for all temperatures below $T_c$ where the specific heat $\tilde{c}$ can be evaluated at either constant volume or pressure.

A major question concerns the behavior of the second sound velocity at intermediate temperatures in dilute
gases. In superfluid Fermi gases with resonant interactions the thermal
expansion coefficient is still relatively small and the formula (\ref%
{c2constantP}) for the second sound velocity turns out to be accurate in a wide
temperature interval. At the same time the thermal expansion is still large
enough to show up in a detectable signal in the density fluctuations
associated with the propagation of second sound. Thanks to this coupling it
has been actually possible to measure the velocity of second sound in Fermi
gases interacting with large values of the scattering length in highly
elongated configurations and to extract the temperature dependence of the
superfluid density\cite{Nature,Hua2013}.

Weakly-interacting Bose gases behave in a very different way due to their large isothermal compressibility. An important consequence of this different behavior is the occurrence of a hybridization phenomenon between first and second sound~\cite{LeeYang, GZ, Griffin2009, Hua2010}. The mechanism is caused by the tendency of the velocity of the two modes to cross at very low temperatures, of the order of the $T=0$ value of the chemical potential $\mu(T=0) = gn$ where $g=4\pi\hbar^2a/m$ is the bosonic coupling constant, with the consequent emergence of a hybridization mechanism. For temperatures below the hybridization point the velocity of the upper branch approaches the Bogoliubov value $c_{B}=\sqrt{gn/m}$, while the lower branch approaches the Landau's result $c_{2}=\sqrt{gn/3m}$. Above the hybridization point the roles of first and second sound are inverted, in the sense that the second sound is essentially an oscillation of the superfluid density which, for temperatures significantly smaller than the critical temperature, coincides with the density of the gas.

The main purpose of the present work is to exploit in an explicit form the mechanism of hybridization, providing an exact expression for the hybridization temperature in the case of weakly-interacting Bose gases and to discuss the behavior of second sound both below and above the hybridization point.
For this purpose we will use Bogoliubov theory to calculate the relevant thermodynamic functions entering Eq.~(\ref{c4}) at low temperatures ($T \ll T_c$), in terms of the elementary excitations of the gas, whose dispersion is given by the famous Bogoliubov law ${\varepsilon(\mathbf{p})=\sqrt{p^2/2m\left(p^2/2m+2gn\right)}}$ and which are thermally excited according to the bosonic rule ${N_\mathbf{p}(\varepsilon)=(e^{\varepsilon(\mathbf{p})/kT} - 1)^{-1}}$. Instead, at higher temperatures we will use the perturbation theory developed in Ref.~\cite{Capogrosso} based on the Beliaev diagrammatic technique at finite temperature~\cite{Abrikosov}. In this approach thermal effects in the Bogoliubov excitation spectrum are accounted for through a self-consistent procedure. At low temperatures this approach coincides with Bogoliubov theory, except for temperatures smaller
than $(na^3)^{1/4}gn$, \emph{i.e.} at temperatures much smaller than the hybridization point. At higher temperatures the theory of Ref.~\cite{Capogrosso} turns out to be very accurate in dilute gases when
compared with exact Monte-Carlo simulations. It follows that, at least for small values of the gas parameter $na^3$, Bogoliubov theory and the diagrammatic approach of Ref.~\cite{Capogrosso} match exactly in the hybridization region of temperatures $kT \sim gn$ and that the thermodynamic behavior of the gas is consequently under control for all ranges of temperatures, both below and above $gn$.

The coefficients of the quartic equation (\ref{c4}) depend not only on the equilibrium thermodynamic functions, but also on the normal and superfluid densities. The normal density can be calculated using the Landau's prescription
\begin{equation}
    \rho_n = - \frac{1}{3} \int \frac{\d N_\mathbf{p}(\varepsilon)}{\d \varepsilon} p^2 \frac{\d \mathbf{p}}{(2\pi \hbar)^3}
    \label{Landau}
\end{equation}
in terms of the elementary excitations of the gas. The Landau's prescription (\ref{Landau}) ignores interaction effects among elementary excitations and in a dilute Bose gas is expected to be very accurate except for temperatures close to the critical point.

A peculiar property of the weakly-interacting Bose gas is that all the thermodynamic functions entering Eq.~(\ref{c4}), as well as the normal density $\rho_n$, can be written in a rescaled form as a function of the reduced temperature $\tilde{t}\equiv kT/gn$ and of the reduced chemical potential $\eta \equiv gn/kT^0_C$ where
\begin{equation}
k T^0_c=\frac{2\pi\hbar^2}{m}\left(\frac{n}{\zeta(3/2)}\right)^{2/3}
 \label{Tc0}
\end{equation}
is the critical temperature of the ideal Bose gas. In weakly-interacting Bose gases $T^0_c$ does not coincide with the actual critical temperature which contains a small correction fixed by the value of the gas parameter $na^3$:
\begin{equation}
    T_c=T_c^0 (1+ \gamma (na^3)^{1/3})
    \label{Tc}
\end{equation}
with $\gamma \sim 1.3$~\cite{Baym, ArnoldMoore, Kashurnikov}. The reduced chemical potential can also be expressed in terms of the gas parameter as $\eta=2\zeta(3/2)^{2/3}(na^3)^{1/3}$.

Using Bogoliubov theory, the free energy $F=U-TS$ can be written as
\begin{equation}
    \begin{aligned}
        \frac{F}{gn N}=&\frac{1}{2}\left[1 + \frac{128}{15\sqrt{\pi}}(na^3)^{1/2}\right]\\
                    &+ \frac{2\tilde{t}}{\zeta(3/2)\sqrt{2\pi}}\eta^{3/2} \int_0^{\infty} \tilde{p}^2 \ln\left(1-e^{-\frac{\tilde{p}}{2\tilde{t}}\sqrt{\tilde{p}^2 + 4}}\right) \d \tilde{p}.\\
        \label{F}
    \end{aligned}
\end{equation}
where we have defined $\tilde{p}=p/\sqrt{mgn}$ and, for completeness, we have included the Lee-Huang-Yang correction to the $t=0$ value of the free energy (term proportional to $(na^3)^{1/2}$).

The thermodynamic functions entering Eq.~(\ref{c4}) are then easily calculated using the following thermodynamic relations ($\rho=mn$)
\begin{equation}
    \tilde{s} = -\frac{1}{m}\left.\frac{\partial F/N}{\partial T}\right|_\rho
\end{equation}

\begin{equation}
    \tilde{c}_v = \frac{1}{m} \left.\frac{\partial (F/N + mT\tilde{s})}{\partial T}\right|_\rho
\end{equation}

\begin{equation}
   \left.\frac{\partial P}{\partial \rho}\right|_{\tilde{s}} = \left.\frac{\partial P}{\partial \rho}\right|_T + \frac{kT}{\rho^2 \tilde{c}_v}\left(\left.\frac{\partial P}{\partial T}\right|_\rho\right)^2
\end{equation}
where $P = -(\partial F / \partial V)_T$ is the pressure of the gas.

The normal density (\ref{Landau}) instead takes the form
\begin{equation}
    \frac{\rho_n}{\rho} = \frac{2}{3\zeta(3/2)\tilde{t}\sqrt{2\pi}}\eta^{3/2} \int_0^{\infty} \frac{\tilde{p}^4 e^{\tilde{p}\sqrt{\tilde{p}^2 + 4}/2\tilde{t}}}{\left(e^{\tilde{p}\sqrt{\tilde{p}^2 + 4}/2\tilde{t}} - 1\right)^2}\d \tilde{p}
    \label{rho_s}
\end{equation}

The idea now is to calculate the two solutions of Eq.~(\ref{c4}) for a fixed value of $\tilde{t}$ of the order of unity, taking the limit $\eta \to 0$. Physically this corresponds to considering very low temperatures (of the order of $gn$) and very small values of the gas parameter $na^3$. For example in the case of $^{87}$Rb the value of the scattering length is $a=100 a_0$ (where $a_0$ is the Bohr radius) and typical values of the density correspond to $na^3 \sim 10^{-6}$. This corresponds to $\eta \approx 0.04$\footnote{Notice that such a value is significantly smaller than the ratio $\mu(T=0)/kT_c$ evaluated in the presence of harmonic trapping with the same value of the gas parameter $na^3$ where is $n$ is the density in the center of the trap~\cite{LevSandro}.}.

Writing the solutions of Eq.~(\ref{c4}) in terms of the $T=0$ Bogoliubov velocity $c_B = \sqrt{gn/m}$, one finds that, as $\eta \rightarrow 0$, the two sound velocities only depend on the dimensionless parameter $\tilde{t}$ and are given by
\begin{equation}
    \left\{\begin{array}{l}
            c^2_+ = c^2_B\\
            c^2_- = c^2_B f(\tilde{t})
    \end{array}\right.
    \label{c+-}
\end{equation}
where $f(\tilde{t}) = \lim_{\eta\rightarrow 0} \frac{\rho T \tilde{s}^2}{\rho_n \tilde{c}_v}\frac{m}{gn}$. Using Eqs.~(\ref{F}-\ref{rho_s}), one actually finds that $\tilde{s} \propto \eta^{3/2}$, $\tilde{c}_v \propto \eta^{3/2}$ and $\rho_n \propto \eta^{3/2}$ and that $f$ is consequently a function of $\tilde{t}$, independent of $\eta$.

\begin{figure}
    \includegraphics[scale=0.65]{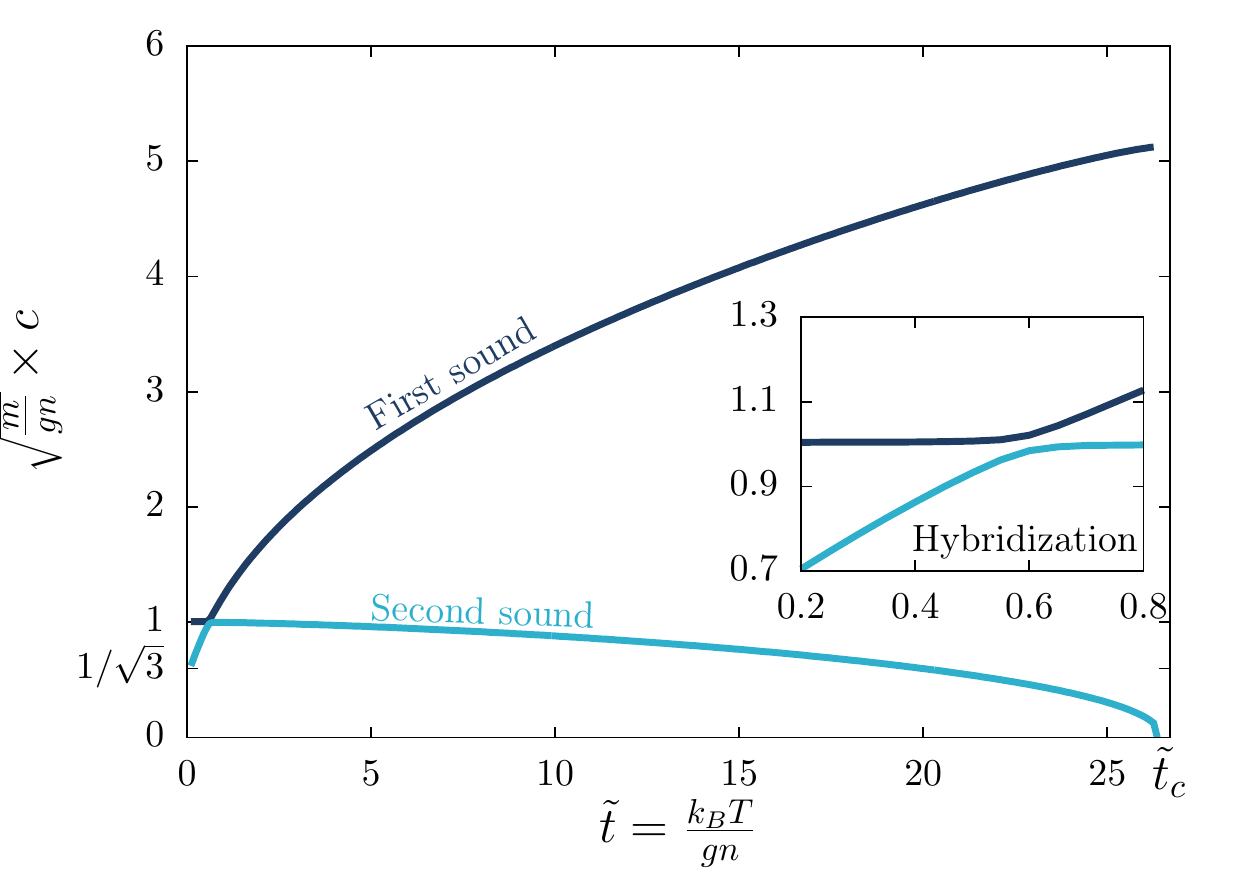}
    \caption{Sound velocities (light and dark blue) computed by interpolating Bogoliubov theory and the diagrammatic approach of Ref.~\cite{Capogrosso}, over the whole range of temperatures $0<\tilde{t}<\tilde{t}_c$. The gas parameter is chosen to be $na^3=10^{-6}$. The critical point (see Eq.~(\ref{Tc})) corresponds to $\tilde{t}_c = k T_c/gn= 26.7$. The inset shows the hybridization region.}
    \label{speeds}
\end{figure}
In the $\eta \rightarrow 0$ limit, the two velocities shown in Fig.~\ref{speeds} cross each other at the value ${\tilde{t}_{\mathrm{hyb}}\approx 0.6}$. At lower temperatures $c^2_-$ approaches, as expected, the zero-temperature value
$c^2_B/3$.

By considering finite, although small, values of $\eta$, it is possible to show that, at the hybridization point, the two branches exhibit a gap proportional to $\eta^{3/4}$. The mechanism of hybridization, explicitly shown in the inset of Fig.~\ref{speeds}, permits to identify an upper branch $c_1$ (which coincides with $c_+$ for $\tilde{t} < \tilde{t}_{\mathrm{hyb}}$ and with $c_-$ for $\tilde{t} > \tilde{t}_{\mathrm{hyb}}$) called ``first sound''. The lower branch $c_2$ (called ``second sound'') instead coincides with $c_-$ for $\tilde{t} < \tilde{t}_{\mathrm{hyb}}$ and with $c_+$ for $\tilde{t} > \tilde{t}_{\mathrm{hyb}}$.

The validity of Eqs.~(\ref{c+-}) is limited to very low temperatures where the thermal depletion of the condensate can be ignored and Bogoliubov theory can be safely applied. When the temperature starts being comparable with the critical temperature $T_c$ (corresponding to the value $\tilde{t}_c=26.7$ in the case of Fig.~\ref{speeds}, where we have chosen $\eta=0.04$), Bogoliubov theory is no longer applicable. In fact at such temperatures the thermal depletion of the condensate becomes important and the Bogoliubov expression for the dispersion law is inadequate. The superfluid density fraction, calculated according to Eq.~(\ref{rho_s}) with the $T=0$ value of the Bogoliubov dispersion law, vanishes at $T\sim 1.2 T_c$, well above the critical temperature (\ref{Tc}), further revealing the inadequacy of the theory at high temperatures. As anticipated above a better approach to be used at temperatures of the order of the critical value is the diagrammatic approach of Ref.~\cite{Capogrosso} whose predictions for the first and second sound velocities at temperatures higher than the hybridization temperature are reported in Fig.~\ref{speeds} and which properly interpolates with the predictions of Bogoliubov theory near the hybridization point.

For temperatures significantly higher than the hybridization temperature an accurate and elegant expression for the second sound velocity is obtained by evaluating all the quantities entering the quartic equation (\ref{c4}), except the isothermal compressibility and the superfluid density, using the ideal Bose gas model. One can actually show that the adiabatic compressibility, the specific heat at constant volume and the entropy density deviate very little from the ideal Bose gas predictions in a wide interval of temperatures above the hybridization point.
This simplifies significantly the calculation of Equation
(\ref{c4}). In fact in the ideal Bose gas model one finds the simple result
\begin{equation}
\left( \frac{\partial P}{\partial \rho }\right) _{\widetilde{s}%
}+\frac{\rho _{s}T \widetilde{s}^{2}}{\rho _{n}\tilde{C}%
_{v}}= \frac{\rho T \tilde{s}^2}{\rho_n\tilde{C}_v}
\label{thermoIBG}
\end{equation}
for the coefficient of the $c^2$ term. It is now easy to derive the two solutions
satisfying the condition $c_{1}\gg c_{2}$ . To obtain the
larger velocity $c_{1}$ one can neglect the last term in $\left( \ref{c4}%
\right) $. Using the thermodynamic relations of the ideal Bose gas model and identifying the normal density with the thermal density ($\rho_n=mn_T=mN_T/V$) one obtains the prediction
\begin{equation}
c_{1}^{2}={5\over 3}{g_{5/2} \over g_{3/2}} \frac{ k_{B}T}{m}
\label{c1B}
\end{equation}
for the first sound velocity
(Lee and Yang ,1959). To calculate $c_{2}$ we must instead neglect the $c^{4}$ term in $\left( \ref
{c4}\right) $ and using result (\ref{thermoIBG}) one finds the useful result
\begin{equation}
    c^2_2 = \frac{\rho_s}{\rho}\left(\frac{\partial P}{\partial \rho}\right)_T
    \label{c-chiT}
\end{equation}
revealing that the superfluid density and the isothermal compressibility are the crucial parameters determining the value of the second sound velocity of weakly-interacting Bose gases for $T\gg gn$. In the relevant region $10\le \tilde{t} \le 25$ of the reduced temperature, the expression Eq.~(\ref{c-chiT}) for the second sound velocity deviates from the exact solution of the two-fluid hydrodynamic equation reported in Fig.~\ref{speeds} by less than $7$ percents.

\begin{figure}
    \includegraphics[scale=0.65]{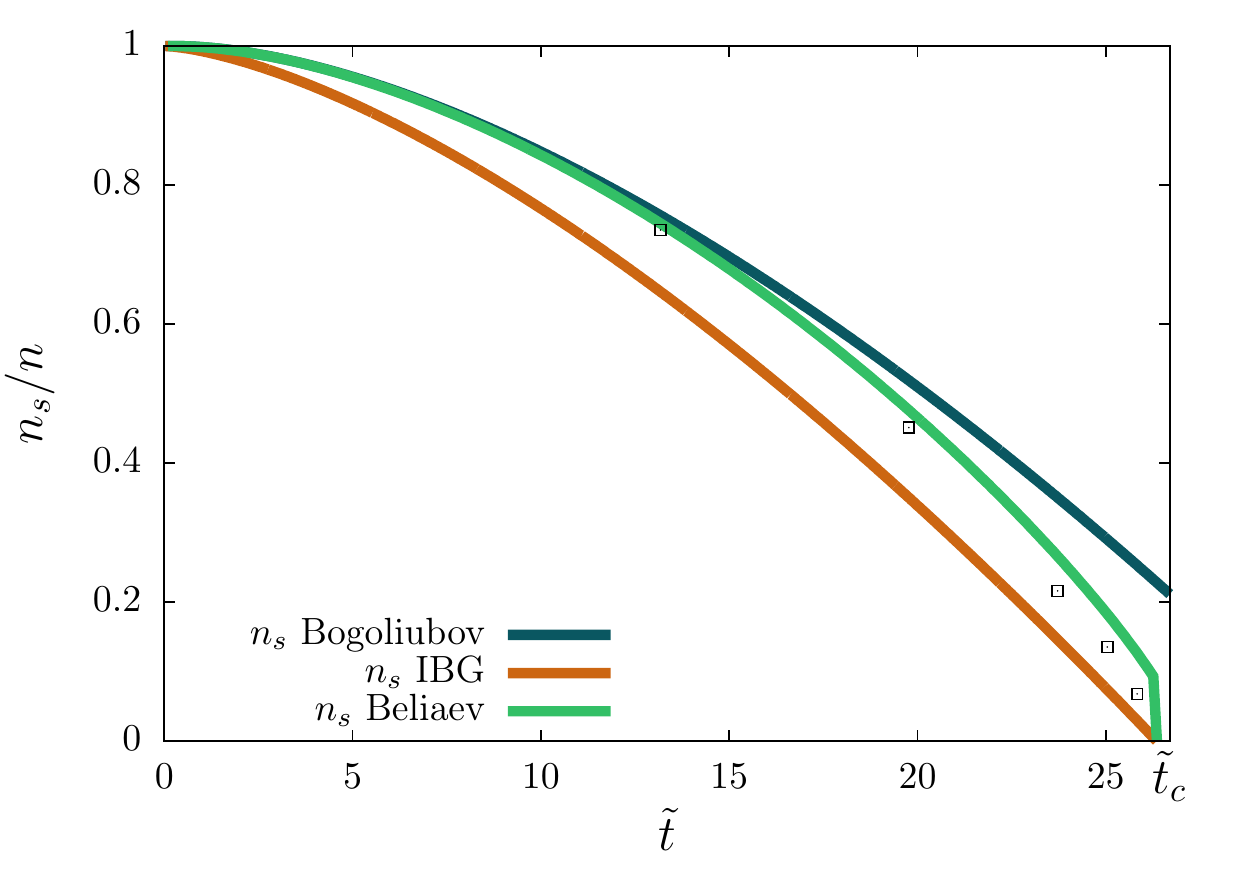}
    \caption{Superfluid density computed using the ideal Bose gas model (orange curve), the Bogoliubov theory (blue curve) and the diagrammatic approach of Ref.~\cite{Capogrosso} (green curve). The black dots correspond to Monte-Carlo simulations. The parameters are chosen as in Fig.~\ref{speeds}.}
    \label{ns_beliaev}
\end{figure}

\begin{figure}
    \includegraphics[scale=0.65]{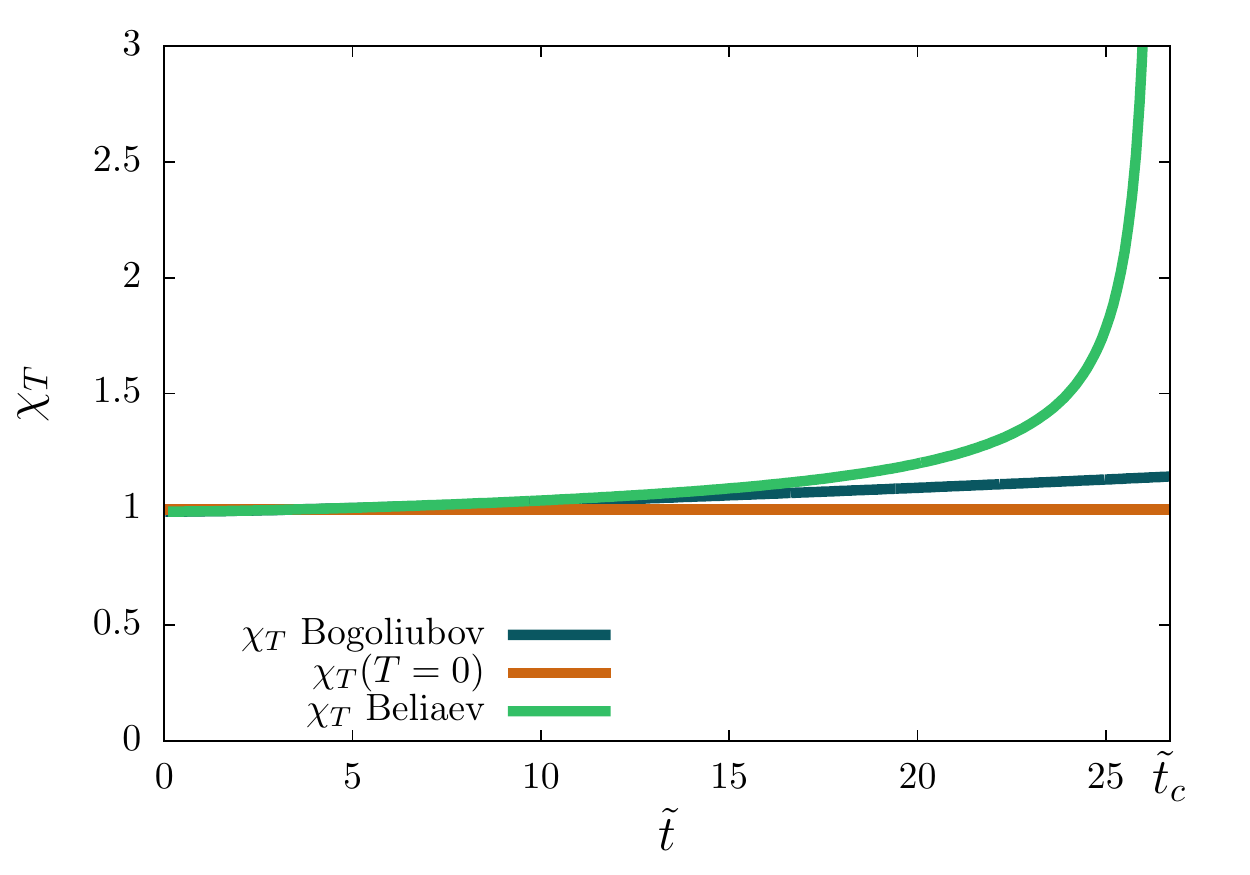}
    \caption{Isothermal compressibility $(\rho \, \partial P / \partial \rho|_T)^{-1}$ computed using the Bogoliubov model (blue curve) and the diagrammatic approach of Ref.~\cite{Capogrosso} (green curve). The parameters are chosen as in Fig.~\ref{speeds}.}
    \label{chiT_beliaev}
\end{figure}

In Fig.~\ref{ns_beliaev} we show the temperature dependence of the superfluid density predicted by the diagrammatic approach of Ref.~\cite{Capogrosso} compared with the predictions of Bogoliubov theory, the results of quantum Monte Carlo simulations as well as the ideal Bose gas value $\rho(1-(T/T_c^0)^{3/2})$. Notice that the diagrammatic approach of Ref.~\cite{Capogrosso} predicts an unphysical jump of the superfluid density at the transition point and that critical fluctuation terms should be taken into account for a correct description near $T_c$. It is also worth discussing the behavior of the isothermal compressibility (see Fig.~\ref{chiT_beliaev}), a quantity of high interest near phase transitions. The diagrammatic approach of Ref.~\cite{Capogrosso}, which at high temperatures approaches Hartree-Fock theory, predicts an increase of the isothermal compressibility $(\rho\, \partial P / \partial \rho|_T)^{-1}$ with respect to its $T=0$ value $(gn)^{-1}$ and a divergent behavior near the critical temperature, which is typical feature of mean field theories. The inclusion of critical fluctuations will modify this behavior near the critical point. To our knowledge the isothermal compressibility near the critical point of dilute Bose gases has not yet been accurately calculated nor measured. This contrasts with resonant Fermi gases where its recent measurement~\cite{zwierlein} was actually employed to identify experimentally the value of the critical temperature.

\begin{figure}
    \includegraphics[scale=0.65]{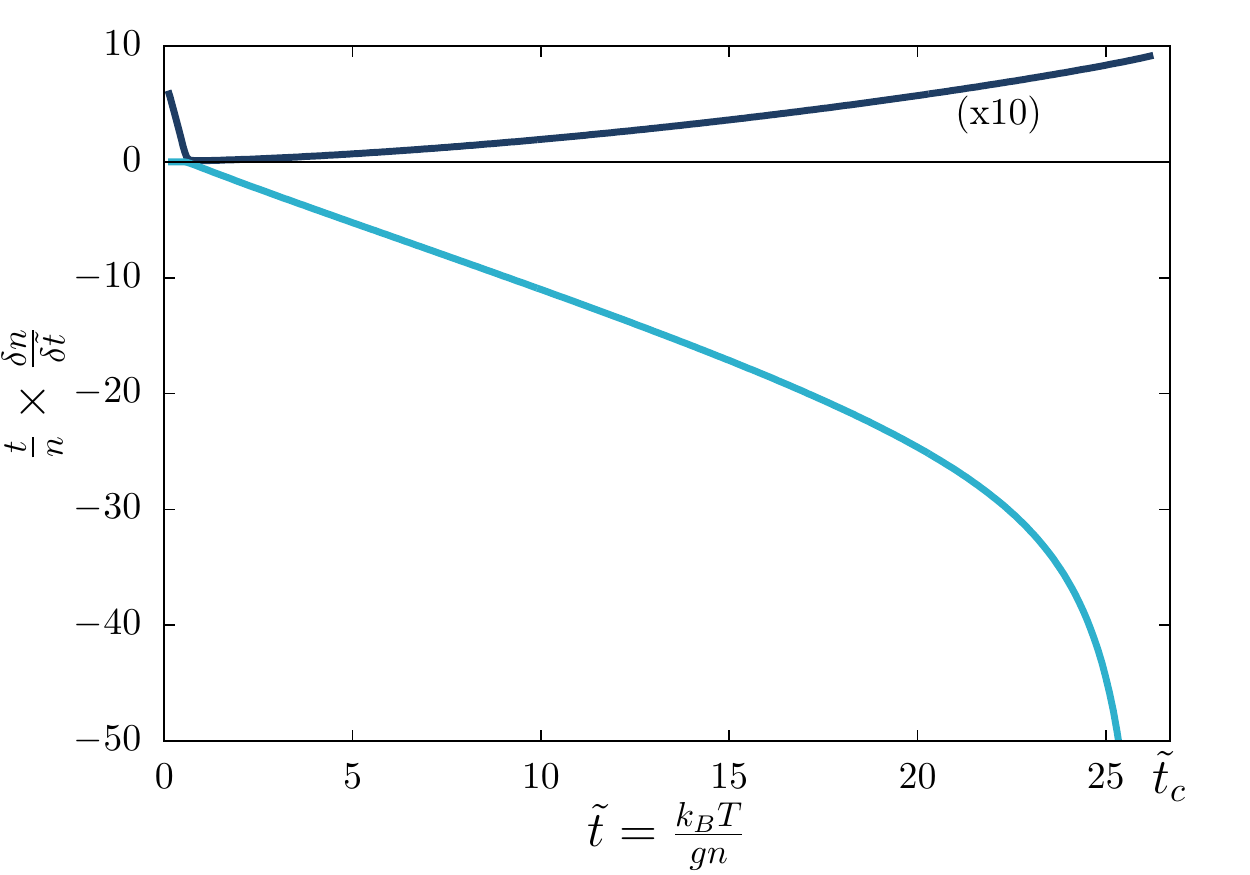}
    \caption{Ratio $(\delta n/n) \times (\tilde{t}/\delta \tilde{t})$ for the lower branch (light blue) and the upper branch (dark blue). The parameters are chosen as in Fig.~\ref{speeds}.}
    \label{delta_n_delta_t}
\end{figure}

In Fig.~\ref{delta_n_delta_t}, we finally show the ratio $(\delta n/n)/(\delta T/ T)$ between the relative density and temperature variations calculated for the first and second sound solutions of Eq.~(\ref{c4}) as a function of temperature. This quantity represents an important characterization of the two branches. It is in fact well known that sound in an ideal classical gas is an adiabatic oscillation characterized by the value $3/2$ for the ratio $(\delta n/n)/(\delta T/ T)$. For the upper solution of the hydrodynamic equations (first sound), this ratio is positive over the full range of temperatures below the transition and its value increases with $T$, getting close to unity near the transition. The most important feature emerging from Fig.~\ref{delta_n_delta_t} is that the ratio between the relative density and temperature changes associated with the second sound solution, have an opposite sign and a much larger value in modulus, reflecting that second sound, for temperatures larger than the hybridization value is dominated by the fluctuations of the density, rather than by the ones of the temperature. This important feature is also revealed by the fact that, in the same range of temperatures, second sound practically exhausts the compressibility sum rule~\cite{LevSandro}
\begin{equation}
(\rho\, \partial P/\partial \rho|_T)^{-1}=\frac{2}{N}\int d\omega S(q,\omega)/\omega \; ,
\label{Sqomega}
\end{equation}
 where $S(q, \omega)$ is the dynamic structure factor, a quantity which can be easily calculated within the two-fluid model described by Landau's theory~\cite{Hu2010}. The fact that the compressibility sum rule (\ref{Sqomega}) is strongly affected by second sound is a remarkable feature exhibited by the weakly-interacting Bose gas above the hybridization point which distinguishes it in a profound way with respect to the case of strongly interacting superfluids, like $^4$He or the Fermi gas at unitarity, where the density fluctuations associated with second sound are instead always very small.

From an experimental point of view this result also reveals that second sound is much more easily accessible than first sound in dilute Bose gases, being very sensitive to the coupling with a density probe, as confirmed by its experimental identification in the experiment of Ref.~\cite{Meppelink}. It is finally worth noticing that the fact that second sound in a Bose gas can be easily excited through a density probe is not specific to the 3D case. Indeed, a similar behavior takes place also in 2D Bose gases~\cite{Tomoki2013} where its measurement could provide an efficient determination of the superfluid density, including its discontinuity at the Berezinski-Kosterlitz-Thouless transition. \revision{Probing directly the hybridization mechanism between first and second sound is a more difficult experimental task because of the extremely low temperatures required to observe the coupling between the two modes. The achievement of the hybridization could be detected by the observation of two sound waves excited by the same density perturbation and propagating with relatively close, but distinct velocities. An alternative perspective is provided by the study of the hybridization exhibited by the frequencies of the discretized modes of a harmonically trapped gas at finite temperature, as discussed in \cite{Griffin2009}.}

Insightful discussions with Stefano Giorgini and D. J. Papoular are acknowledged. L.V. would like to acknowledge the ICFP undergraduate program of the \'{E}cole normale sup\'{e}rieure and the hospitality of the Trento Physics Department. This work was supported by the ERC through the QGBE grant and by Provincia Autonoma di Trento.

\bibliographystyle{eplbib}
\bibliography{lucas}
\end{document}